\documentclass[aps,prl,twocolumn,amsmath,amssymb,floatfix,showkeys,showpacs]{revtex4-1}
\usepackage{bm} % bold math
\usepackage{color}
\usepackage{graphics} %   <------- for LatTeX + PS
\usepackage{graphicx} % Include figure files
\usepackage{epsfig}
\usepackage{amssymb}
\usepackage{amsmath} %,amssymb}
\usepackage{amsthm}
\usepackage{amstext}
\usepackage{dcolumn} % Align table columns on decimal point
\usepackage{makeidx}
\usepackage{subfigure}

\begin{document}

\title{Dissipative two-level systems under ultrastrong off-resonant driving}

\author{A.P. Saiko}
\email{saiko@physics.by}
\affiliation{Scientific-Practical Material Research Centre, Belarus National Academy of Sciences, 19 P. Brovka str., Minsk 220072 Belarus}

\author{R. Fedaruk}
\affiliation{Institute of Physics, University of Szczecin, 15 Wielkopolska str., 70-451, Szczecin, Poland}

\author{S.A. Markevich}
\affiliation{Scientific-Practical Material Research Centre, Belarus National Academy of Sciences, 19 P. Brovka str., Minsk 220072 Belarus}

\date{\today}

\begin{abstract}
We study the dissipative dynamics of a two-level system under ultrastrong driving when the frequency and strength of the exciting field exceed significantly the transition frequency. We find three qualitatively different regimes of such dynamics: 1) the  collapse and revival of oscillations in the population difference, 2) the simple exponential decay of the oscillations resulting in their steady state with the finite amplitude, and 3) the steady-state stabilization of the equally populated levels. The nonmonotonic Bessel-function-like dependence on the driving strength is also predicted for the decay rate of these oscillations. The features of this dependence are determined by the relative rates of energetic relaxation and pure dephasing.
\end{abstract}


\maketitle
\section{I. INTRODUCTION}
The resonant interaction between electromagnetic fields and two-level systems (qubits) is widely used for studying and control of various quantum objects such as atoms, nuclear and electronic spins, impurity centers, quantum dots, and superconducting qubits \cite{pp1, pp2, pp3}. The coherent dynamics of qubits can be described in terms of Rabi oscillations between the two energy eigenstates. In particular, this dynamics is extremely important for quantum information processing \cite{pp41}, quantum control \cite{pp4}, and protection against decoherence \cite{pp42}. The rate of the two-level state manipulation and the system coherence time are critical parameters for the processing. The manipulation rate is characterized by the Rabi frequency and depends on the strength of the driving field. The number of coherent single-qubit operations (the number of half-periods of Rabi oscillations) is limited by the system coherence time. Both the lengthering of the coherence time and the increase in the manipulation rate result in faster state operation. Increasing the manipulation rate requires stronger driving fields and can lead to the strong driving regime when the driving strength $g$ is comparable to, or exceeds, the transition frequency $\varepsilon $ between two energy levels. In this regime the counter-rotating component of an oscillatory driving field results in complex dynamics of two-level systems due to breakdown of the rotating wave approximation (see, e. g., \cite{pp5}). Previous results on the steady-state response of two-level systems, mainly superconducting qubits, under their strong continuous-wave driving have been reviewed  \cite{pp2, pp6}. Recently, using time-domain Rabi oscillations, the strong driving ($g\ge \varepsilon $) has been studied in experiments with nuclear spins \cite{pp7}, artificial atoms such as superconducting flux \cite{pp8, pp9, pp10} and charge \cite{pp11} qubits, a single nitrogen-vacancy (NV) center in diamond \cite{pp12, pp13}, radiation-dressed states of NV centers \cite{pp14}, and mechanical driving of a single electron spin  \cite{pp15}. Usually the qubit's dynamics under strong driving is described within the framework of Floquet theory, where the state of a driven system is expressed in terms of quasienergies and quasienergy states \cite{pp10}. The presence of the various frequency components in the observed Rabi oscillations and the Bessel-function dependence of the quasienergy difference on the driving strength have been demonstrated  \cite{pp9, pp10, pp11}. On the other hand, the study of dissipative and decoherence processes limiting the observation of Rabi oscillations remains a challenging task in the strong and ultrastrong ($g>>\varepsilon $) driving regime.

In this paper, we present an analytical description of the dissipative dynamics under the ultrastrong driving, when the frequency $\omega $ and the strength of electromagnetic field exceed significantly the qubit transition frequency ($\omega, g>>\varepsilon $). The description is given in the framework of the non-secular perturbation theory based on the Krylov--Bogoliubov--Mitropolsky (KBM) averaging method. With the help of the KBM method the secular terms in the series of the perturbation theory are resummed and the accurate results up to the third order in $\sim \varepsilon /\omega $ are obtained. We predict three regimes of oscillations in the population difference of the qubit. These regimes with qualitatively different behavior are controlled by the driving strength. The unusual feature of these oscillations is also the nonmonotonic Bessel-function-like dependence of their decay rate on the driving strength. Such properties cannot be realized under weak driving, where the decay rate of Rabi oscillations  can increase monotonically with the driving strength, an effect called ``driven decoherence'' \cite{pp16}. The obtained analytical results are validated by numerical calculations.

\section{II. DENSITY-MATRIX EVOLUTION}
The master equation for the qubit interacting with a linearly polarized electromagnetic field is
\begin{equation} \label{eq1}
i\frac{\partial \rho }{\partial t} =\left[H,\rho \right]+i\Lambda \rho ,
\end{equation}
where
\begin{equation} \label{eq1_1}
H=\varepsilon s^{z} -2gs^{x} \cos \omega t
\end{equation}
is the Hamiltonian and $\Lambda $ is the relaxation superoperator defined as
\begin{equation} \label{eq1}
\Lambda \rho =\frac{\gamma _{21} }{2} D[s^{-} ]\rho +\frac{\gamma _{12} }{2} D[s^{+} ]\rho +\frac{\eta }{2} D[s^{z} ]\rho.
\end{equation}
 Here $g$ is the qubit-field coupling strength, $s^{\pm ,\, z} $ are components of the pseudospin operator, describing the qubit state and satisfying the commutation relations: $[s^{+} ,s^{-} ]=2s^{z} $, $[s^{z} ,s^{\pm } ]=\pm s^{\pm } $. In addition, $D[O]\rho =2O\rho O^{+} -O^{+} O\rho -\rho O^{+} O$, $\gamma _{21} $ and $\gamma _{12} $ are the rates of photon radiative processes from the excited state ${\left| 2 \right\rangle} $ of the qubit to its ground state ${\left| 1 \right\rangle} $ and vice versa, and $\eta $ is the dephasing rate. Note that the strong interaction between the external field and the qubit should be taken into account at the microscopic definition of the relaxation superoperator, as it was done, e.g., in \cite{pp17, pp18, pp19, pp191}. However, for our purposes, it is sufficient to use the superoperator in its standard form, assuming that the relaxation parameters are defined phenomenologically.

After the canonical transformation $\rho _{1} =u^{+} \rho u$, where $u=\exp (i\frac{2g}{\omega } s^{x} \sin \omega t)$, the master equation is transformed into

\begin{equation} \label{eq2}
i\frac{\partial \rho _{1}}{\partial t} =\left[H_{1} ,\rho _{1} \right]+i\Lambda _{1} \rho _{1} ,
\end{equation}
\begin{equation} \label{eq2_1}
H_{1} =u^{+} Hu-iu^{+} \frac{\partial u}{\partial t} =\frac{\varepsilon}{2}\left[s^{z} +(s^{+} -s^{-} )/2\right]f(t)+h.c.,
\end{equation}

\begin{equation} \label{eq2_2}
\Lambda _{1} \rho _{1} =u^{+} \Lambda u\rho _{1} =\frac{\Gamma _{\downarrow } }{2} D[s^{-} ]\rho _{1} +\frac{\Gamma _{\uparrow } }{2} D[s^{+} ]\rho _{1} +\frac{\Gamma _{\varphi } }{2} D[s^{z} ]\rho _{1} ,
\end{equation}
where
\begin{equation} \label{eq2_3}
\begin{split}
\Gamma _{\downarrow } =(\gamma _{21} a_{+}^{2} +\gamma _{12} a_{-}^{2} -\eta c^{2} /4)/2,\\
\Gamma _{\uparrow } =(\gamma _{12} a_{+}^{2} +\gamma _{21} a_{-}^{2} -\eta c^{2} /4)/2,\\
\Gamma _{\varphi } =\left(\eta d^{2} -(\gamma _{12} +\gamma _{21} )c^{2} \right)/2;
\end{split}
\end{equation}
$a_{\pm } =\left[1\pm d\right]$, $c=Im\left(f(t)\right)$, $d=Re\left(f(t)\right)$, $f(t)=\exp \left[ia\sin \omega t\right]$, and $a=2g/\omega $.
$$ $$
\section{III. AVERAGING OF RAPID OSCILLATIONS}

The rapidly oscillating terms in the transformed master equation  can be eliminated in the framework of the non-secular perturbation theory by using the KBM averaging method \cite{pp20}. The description of this method and its applications to studies of the dynamics of two-level systems under bichromatic driving  \cite{pp201, pp21}  as well as strongly correlated electron systems in solid state physics \cite{pp202} have been published previously. In the high-frequency limit, $\varepsilon /\omega <<1$, the Hamiltonian $H_{1} $  is replaced by its effective counterpart up to the third order in this small parameter: $H_{1} \to H_{eff}= H_{eff}^{(1)} +H_{eff}^{(3)} $, where
\begin{equation} \label{eq2_4}
H_{eff}^{(1)} =\left\langle H_{1} \right\rangle =\varepsilon J_{0} (a)s^{z} \equiv \varepsilon ^{(1)} s^{z},
\end{equation}
\begin{widetext}
\begin{equation} \label{eq3}
H_{eff}^{(3)} =-\frac{1}{3} <[\int _{}^{t}d\tau (H_{1} (\tau )-<H_{1} (\tau )>) ,[\int _{}^{t}d\tau (H_{1} (\tau )-<H_{1} (\tau )>) ,(H_{1} (t)+\frac{1}{2} <H_{1} (t)>)]]>=\varepsilon ^{(3)} s^{z},
\end{equation}
\[\varepsilon ^{(3)}=\frac{\varepsilon ^{3} }{4\omega^{2}} \left\{J_{0} (a)\sum _{n\ne 0}\frac{J_{n}^{2} (a)}{n^{2} } \left((-1)^{n} -1\right)-\frac{2}{3} \left[\sum _{\substack {n\ne 0, n_{1} \ne 0, \\ n+n_{1} \ne 0 }}\frac{J_{n} (a)J_{n_{1} } (a)J_{n+n_{1} } (a)}{nn_{1} } +\sum _{\substack{n \ne 0, n_{1} \ne 0, \\ n-n_{1} \ne 0 }}\frac{J_{n} (a)J_{n_{1} } (a)J_{n_{1} -n} (a)}{nn_{1} }\right]\right\}.\]

\end{widetext}
Here the symbol $\langle ...\rangle $ denotes time averaging over the period $2\pi /\omega $ of the rapid oscillations $\exp (in\omega t)$ given by $\langle O(t)\rangle =\frac{\omega }{2\pi } \int _{0}^{{2\pi \mathord{\left/ {\vphantom {2\pi  \omega }} \right. \kern-\nulldelimiterspace} \omega } }O(t)dt $, where $n=\pm 1,\pm 2,...$, and $O(t)$ is the some time-dependent operator. Square brackets in the definition of the effective Hamiltonians $H_{eff}^{(1,3)} $ denote the commutation operation and the upper limit \textit{t} of the indefinite integral indicates the variable on which the result of the integration depends. $J_{n} $ is the Bessel function of the first kind and order \textit{n}. The second order of the nonsecular perturbation theory does not yield the contribution in the effective Hamiltonian. The relaxation operator calculated in the first non-vanishing approximation is given by
\begin{equation} \label{eq3_1}
\begin{split}
\left\langle \Lambda _{1} \right\rangle \left\langle \rho _{1} \right\rangle =\frac{\left\langle \Gamma _{\downarrow } \right\rangle }{2} D[s^{-} ]\left\langle \rho _{1} \right\rangle+\\
 +\frac{\left\langle \Gamma _{\uparrow } \right\rangle }{2} D[s^{+} ]\left\langle \rho _{1} \right\rangle +\frac{\left\langle \Gamma _{\varphi } \right\rangle }{2} D[s^{z} ]\left\langle \rho _{1} \right\rangle,
\end{split}
\end{equation}
where
\begin{equation} \label{eq3_2}
\begin{split}
\left\langle \Gamma _{\downarrow } \right\rangle =\frac{\gamma }{8} \left(3+4J_{0} (a)+J_{0} (2a)\right)-\frac{\eta }{8} \left(J_{0} (2a)-1\right),\\
  \left\langle \Gamma _{\uparrow } \right\rangle =\frac{\gamma }{8} \left(3-4J_{0} (a)+J_{0} (2a)\right)-\frac{\eta }{8} \left(J_{0} (2a)-1\right),\\
  \left\langle \Gamma _{\varphi } \right\rangle =\frac{\gamma +\eta }{2} +\frac{\eta -\gamma }{2} J_{0} (2a).
\end{split}
\end{equation}

 In the above equations, we denoted $\gamma _{21}$ by $\gamma $ and assumed that at low temperatures $\gamma _{12} \approx 0$. Then, we obtain:
  \begin{equation} \label{eq3_4}
i\frac{\partial \left\langle \rho _{1} \right\rangle} {\partial t}=\left[H_{eff}^{} ,\left\langle \rho _{1} \right\rangle \right]+i\left\langle \Lambda _{1} \right\rangle \left\langle \rho _{1} \right\rangle .
\end{equation}

Within the first-order approximation the quasienergies $\pm \varepsilon ^{(1)} /2=\pm \varepsilon J_{0} (a)/2$ are equal to zero for such driving strengths for which the Bessel function $J_{0} (a)$ is equal to zero (Fig.~\ref{fig1}) \cite{pp22}. The third-order correction to the quasienergies consists of two parts. The first one (the first sum in $H_{eff}^{(3)} $) is proportional to $J_{0} (a)$. The second one (the residual two sums in Eq. (\ref{eq3})) is very small but, due to its contribution, the zeroth values of the quasienergies $\pm (\varepsilon ^{(1)} +\varepsilon ^{(3)} )/2$ are shifted from the zeros of the Bessel function $J_{0} (a)$ within the range that is smaller than 0.01\textit{a}, as it is shown in the insets in Fig.~\ref{fig1} (see also \cite{pp23,pp24}. Assuming that the Hamiltonian $H$ describes the tunneling of a particle in a double-well potential at the action of a sinusoidal exciting field, the crossings of quasienergies correspond to the blocking of tunneling dynamics of the system, i.e. the effect termed ``coherent destruction of tunneling'' \cite{pp17, pp25} is realized. The problem of calculations of the third-order correction to the quasienergies in the high-frequency limit has been prevoiously discussed  \cite{pp24}. Different approaches such as the averaging method \cite{pp23}, the dual Dyson series and renormalization group techniques \cite{pp24} give different expressions with the sums of multiplications of the Bessel functions of all orders. It is difficult to reduce these expressions to each other but the third-order corrections obtained numerically by using these approaches differ insignificantly. Our approach based on the non-secular perturbation theory with the KBM averaging method gives the results consistent with those obtained by the previous methods.

Taking into account that
\begin{equation} \label{eq3_2}
\begin{split}
e^{(-iL_{eff} +\left\langle \Lambda _{1} \right\rangle )t} s^{\pm } =\\
=e^{\left(\mp i(\varepsilon ^{(1)} +\varepsilon ^{(3)} )-\left\langle \Gamma _{\bot } \right\rangle \right)t} s^{\pm } ,\\
 e^{(-iL_{eff} +\left\langle \Lambda _{1} \right\rangle )t} s^{z} =e^{-\left\langle \Gamma _{||} \right\rangle t} s^{z} ,\\ e^{(-iL_{eff} +\left\langle \Lambda _{1} \right\rangle )t} const=\\
 =[1+2\sigma _{0} (1-e^{-\left\langle \Gamma _{||} \right\rangle t} )s^{z} ]const,\\
\end{split}
\end{equation}
\begin{equation} \label{eq3_3}
\rho (0)={1\mathord{\left/ {\vphantom {1 2-s^{z} }} \right. \kern-\nulldelimiterspace} 2-s^{z} }
\end{equation}
 and that the superoperator $L_{eff} $ acts in an accordance with the rule:  $L_{eff} X=[H_{eff} ,X],$
 the density matrix in the laboratory frame can be written as:
\begin{equation} \label{eq4}
 \begin{split}
\rho (t)={1}/{2} +\left(\sigma _{0} -(\sigma _{0} +1)e^{-\left\langle \Gamma _{||} \right\rangle t} \right)\times \\
\times\left(\cos (a\sin \omega t)s^{z} -\frac{i}{2} \sin (a\sin \omega t)(s^{+} -s^{-} )\right),
\end{split}
\end{equation}
where
\begin{equation} \label{eq4_1}
\sigma _{0} =-(\left\langle \Gamma _{\downarrow } \right\rangle -\left\langle \Gamma _{\uparrow } \right\rangle )/\left\langle \Gamma _{||} \right\rangle,
\end{equation}
\begin{equation} \label{eq5}
\begin{split}
\left\langle \Gamma _{||} \right\rangle =\left\langle \Gamma _{\downarrow } \right\rangle +\left\langle \Gamma _{\uparrow } \right\rangle,\\
\left\langle \Gamma _{\bot } \right\rangle =(\left\langle \Gamma _{\downarrow } \right\rangle +\left\langle \Gamma _{\uparrow } \right\rangle +\left\langle \Gamma _{\varphi } \right\rangle )/2,\\
\left\langle \Gamma _{||} \right\rangle =\frac{3}{4} \gamma +\frac{1}{4} \eta +\frac{1}{4} (\gamma -\eta )J_{0} (2a),\\
\left\langle \Gamma _{\bot } \right\rangle =\frac{5}{8} \gamma +\frac{3}{8} \eta +\frac{1}{8} (\eta -\gamma )J_{0} (2a).
\end{split}
\end{equation}
We find the population difference of the initial qubit
\begin{equation} \label{eq6}
W=Sp\left(s^{z} \rho (t)\right)=\Xi (t)\cos \left(a\sin \omega t\right),
\end{equation}
where
\begin{equation} \label{eq6_1}
\Xi (t)=\left[\sigma _{0} -(\sigma _{0} +1)e^{-\left\langle \Gamma _{||} \right\rangle t} \right]/2
\end{equation}
 is the relaxation coefficient, $\sigma _{0} =-\gamma J_{0} (a)/\left\langle \Gamma _{||} \right\rangle $ in accordance to \eqref{eq4_1}. When the relaxation is ignored ($\left\langle \Gamma _{||} \right\rangle =0$), this expression coincides with the one obtained for the strong quantum field  \cite{pp26}.

 Within the approximation considered here ($\omega ,g\gg\varepsilon $),   $\varepsilon $ does not enter the expressions for the relaxation parameters and the population difference. In order to applying the KBM method, we must use the values of $a\ge 1.5$.

Physically, the ultrastrong high-frequency ($\omega \gg\varepsilon $) field modulates the energy gap of the qubit and excites parametrically quantum transitions in the coupled qubit-field system. These transitions are realized at frequencies $\varepsilon _{q} +2n\omega $ ($n=0,1,2,3...$), where $\varepsilon _{q} \equiv \varepsilon ^{(1)} +\varepsilon ^{(3)} $ is the quasienergy, and can be observed in the spectrum of resonance scattering. At the same time, the oscillations of the population difference occur at frequencies $2n\omega $ without the frequency component $\varepsilon _{q} $.
\section{IV. REGIMES OF QUBIT'S TIME EVOLUTION}
\begin{figure}[]
\centering
\includegraphics[]{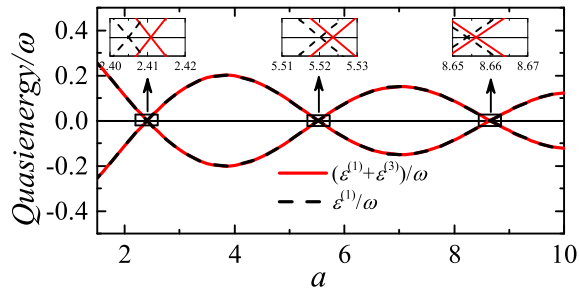}
\caption{Quasienergies versus the driving strength. Solid lines are the first approximation, while dashed lines are obtained with the third-order correction. Insets show the zeroth values of the quasienergies in more detail.}
\label{fig1}
\end{figure}

\begin{figure}[]
\centering
\includegraphics[]{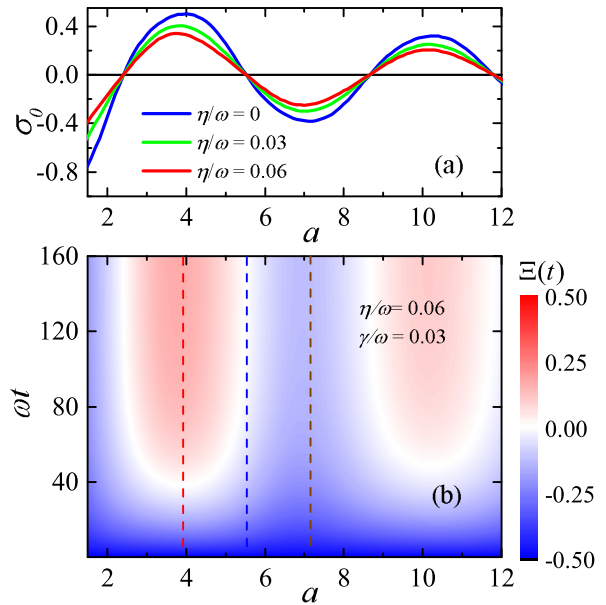}
 \caption{(a) The parameter $\sigma _{0} $ as a function of the driving strength at the energetic relaxation rate $\gamma /\omega =0.03$ and three values of the pure dephasing rate $\eta $. (b) The time evolution of the population difference as a function of the driving strength. The vertical dashed lines show the values of the driving strength for which the time evolution of the population difference is presented in Fig.~\ref{fig3}.}
\label{fig2}
\end{figure}

Now we consider the qubit dynamics in more detail. The parameter $\sigma _{0} $ is equal to twice the population difference of the quasienergy states at $t\to \infty $. Due to the presence of $J_{0} (a)$ in the expression for $\sigma _{0} $, this parameter oscillates and can be negative, zero or positive (Fig.~\ref{fig2}(a)). We observe that the ratio of the rates of energetic relaxation and pure dephasing influence only the amplitude of the variations in $\sigma _{0} $, but  cannot change the sign of $\sigma _{0} $. The variations of the parameter $\sigma _{0} $ correspond to the Bessel-function dependence of the quasienergy states on the driving strength (see Fig.~\ref{fig1}).

\begin{figure}[]
\centering
\includegraphics[]{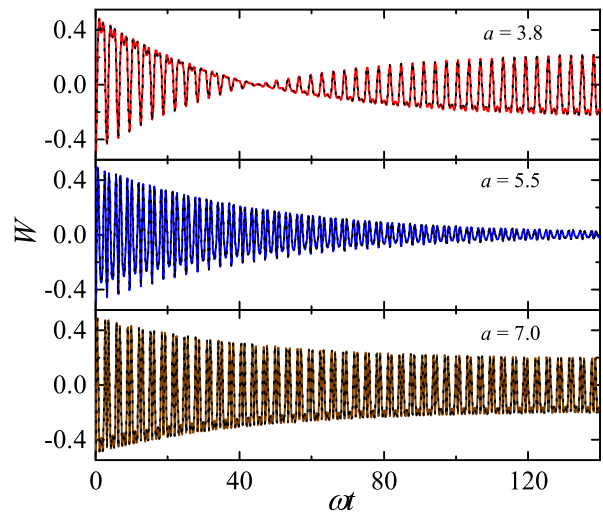}
\caption{ The time evolution of the population difference $W$ for three driving strengths corresponding to three regimes of the oscillations. The collapse and revival of the oscillations is realized at \textit{a }= 3.8. The steady-state stabilization of the oscillations is presented at \textit{a }= 5.5. At \textit{a }= 7.0 the decaying and steady-state oscillations are observed.}
\label{fig3}
\end{figure}

\begin{figure}[]
\centering
\includegraphics[]{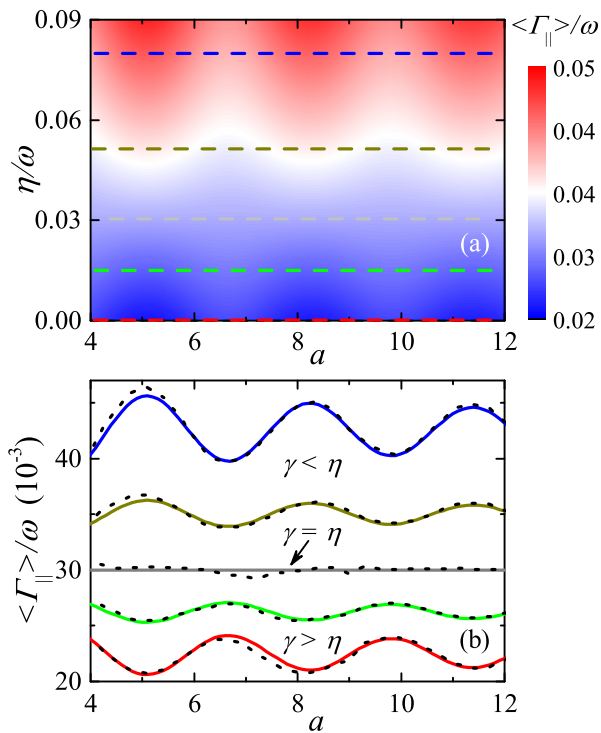}
\caption{(a) The relaxation rate $\left\langle \Gamma _{||} \right\rangle $ of the population difference as a function of the driving strength and the dephasing rate at the energetic relaxation rate $\gamma /\omega =0.03$. (b) Cuts of (a) at $\eta =0$ (red line), $\eta /\omega =0.015$ (green line), $\gamma =\eta $ (grey line), $\eta /\omega =0.05$ (brown line) and $\eta /\omega =0.08$ (blue line). The dashed lines of same colors show positions of these cuts in (a). The dotted lines in (b) present the numerical results.}
\label{fig4}
\end{figure}

The color plot shows the time evolution of the relaxation coefficient  $\Xi (t)$ in the population difference $W$ of the qubit as a function of the driving strength (Fig.~\ref{fig2}(b)). The plot demonstrates three different time dependences of the relaxation coefficient  which result in three possible regimes of the time evolution of the qubit excited from its ground state. These regimes are determined by the population difference of the quasienergy states, which can be negative, positive or zero depending on the driving strength.  As a result, three qualitatively different regimes of the oscillations in the population difference $W$ can be observed. These time-domain oscillations occur at frequencies $2n\omega $ with the intensities proportional to $J_{2n} \left(a\right)$, where $n$ = 1,2,3... In contrast to the strong \textit{resonant} driving \cite{pp10}, there are no oscillations with the frequency of the quasienergy difference. At the strong \textit{off-resonant} driving the quasienergy states determine the regimes of the dissipative dynamics of the qubit.

\section{V. DRIVING-DEPENDENT RELAXATION RATE OF POPULATION DIFFERENCE}
For the values of the driving strength resulting in $\sigma _{0} =0$, the oscillations decay to zero with the characteristic time $\left\langle \Gamma _{||} \right\rangle ^{-1} $. For these values of $a$, the vertical lines are tangent to the white areas in Fig.~\ref{fig2}(b) and the relaxation coefficient  $\Xi (t)$ decay to zero.  In this case, the system relaxes to the steady state with the equally populated levels, i.e. the stabilization of these levels occurs. Fig.~\ref{fig3} shows such oscillations for $a = 5.5$. If the variation of $\Xi (t)$ is within the blue areas in Fig.~\ref{fig2}(b) ($\sigma _{0} <0$), the decaying oscillations
disappear with the characteristic time ${\left\langle {{\Gamma _{||}}} \right\rangle ^{ - 1}}$ and only the steady-state oscillations with the fixed amplitude $\left| {{\sigma _0}} \right|$  remain. Fig.~\ref{fig3} depicts such oscillations for $a$\textit{ }= 7.0. If $\sigma _{0} >0$, $\Xi (t)$ changes its sign at some driving strengths and passes from the blue area to the red one crossing over the white area in Fig.~\ref{fig2}(b). In this case, the oscillations of the population difference decay to zero during the time $t_{c} =\left\langle \Gamma _{||} \right\rangle ^{-1} \ln (1+\sigma _{0}^{-1} )$, then revive with the characteristic time $\left\langle \Gamma _{||} \right\rangle ^{-1} $ and reach their steady-state amplitude $\sigma _{0} $, as it is shown for $a$\textit{ }= 3.8 (Fig.~\ref{fig3}). We observe the collapse and revival of the oscillations. In contract to the well-known collapse and revival of Rabi oscillations in quantum optics \cite{pp27}, this effect is not caused by quantum properties of radiation. It arises due to the competition between the steady-state oscillations with the constant amplitude $\sigma _{0} /2$ and the exponentially decaying oscillations with the amplitude $-e^{-\left\langle \Gamma _{||} \right\rangle t} (\sigma _{0} +1)/2$. Since at $\sigma _{0} <0$ the amplitudes of the steady-state and decaying oscillations have the same sign and at $\sigma _{0} =0$ the oscillations with the amplitude $-e^{-\left\langle \Gamma _{||} \right\rangle t} /2$ only occur, in these cases the collapse and revival effect is absent.

Fig.~\ref{fig4} depicts the qubit relaxation rate $\left\langle \Gamma _{||} \right\rangle $ of the population difference versus the normalized driving strength $a=2g/\omega $ and the pure dephasing rate $\eta $ at the fixed energetic relaxation rate $\gamma $. These dependences were calculated under the assumption that we have capability of changing the coupling of the qubit to its environment, i. e. the relaxation rates. Such changes can be realized by choosing natural quantum systems in corresponding materials or by using artificial atoms. So, for artificial atoms such as semiconductor quantum dots, pure dephasing processes are almost absent \cite{pp28}. On the other hand, the spin coherence time of solid-state qubits such as NV centers in diamonds can be changed over a wide range using its dependence on the concentration of paramagnetic centers \cite{pp29}. Moreover, the dephasing rate can be controlled by means of external stochastic fields modulating the resonant transition frequency of qubits \cite{pp30, pp31, pp32}. We observe in Fig.~\ref{fig4} that the ultrastrong driving significantly modifies not only the energy states of the two-level system, but also its relaxation behavior. Due to such modification, the relaxation rate $\left\langle \Gamma _{||} \right\rangle $ depends on the driving strength in a very unusual way. The features of this dependence are determined by the ratio of the rates of energetic relaxation $\gamma $ and pure dephasing $\eta $. When the pure dephasing is absent ($\eta =0$) or when $\gamma >\eta $, the relaxation rate $\left\langle \Gamma _{||} \right\rangle $ oscillates in accordance with the Bessel-function dependence $J_{0} (2a)$. At $\gamma <\eta $, the variations of $\left\langle \Gamma _{||} \right\rangle $ are inverted in comparison with the previous case because the Bessel function changes its sign. The equality $\gamma =\eta $ is the condition for the crossover between these regimes. Upon satisfaction of this condition the relaxation rate $\left\langle \Gamma _{||} \right\rangle $ is independent of the driving strength and is the same as under weak driving. The decay rate of Rabi oscillations of artificial atoms at weak non-resonant excitation has been investigated in \cite{pp33}. When the condition $\gamma =\eta $ is not fulfilled, the strong driving decreases (at $\gamma >\eta $) or increases (at $\gamma <\eta $) the relaxation rate $\left\langle \Gamma _{||} \right\rangle $ of about half of its value at the weak driving. We see in Fig.~\ref{fig4} that the amplitude of the variations of the relaxation rate increases when the difference between $\gamma $ and $\eta $ increases. For example, at $\gamma =8\eta /3$ the variations of the relaxation rate is of about 12 \% (the blue line in Fig.~\ref{fig4}(b)). In Fig.~\ref{fig4}(b) the analytical and numerical results are presented by the solid and dotted lines, respectively. There is a good agreement between these results.

\section{VI. CONCLUSIONS}
We have studied the dynamics of a qubit under an ultrastrong non-resonant high-frequency driving. The problem was analytically solved in the framework of the non-secular perturbation theory based on the Krylov--Bogoliubov--Mitropolsky averaging method. We have found three qualitatively different regimes in the dissipative dynamics of the qubit. The realization of these regimes is determined by the driving strength. When the driving field inverts the quasienergy states, the collapse and revival of the time-domain oscillations in the population difference of the initial two-level system is observed. At the degeneration of the quasienergy states, the simple exponential vanishing of the oscillations (the steady-state stabilization of equally populated levels) takes place. If the lower and upper quasilevels are  separated enough, the decaying oscillations of the population difference disappear due to relaxation processes and only the steady-state oscillations  remain.  Moreover, we predicted that the ultrastrong off-resonant  driving modifies the decay rate of the oscillations in the population difference and can cause its nonmonotonic Bessel-function-like dependence on the driving strength. Our results are confirmed by numerical calculations.
We expect that our theoretical results will stimulate future experiments to verify our predictions on the dissipative qubit's dynamics under ultrastrong driving. The discovered features of this dynamics are fundamental and important to the physics of open quantum systems as well as for practical applications, including coherent transient spectroscopy and quantum information. In particular, the regime with the steady-state oscillations can be used for long-time manipulations to quantum information because in this regime the coherent oscillations are not limited by the relaxation times or by a rapid collapse.
 Note that the  expression obtained for the density matrix allows one to calculate the coherent response of open quantum systems under their multi-pulse ultrastrong off-resonant excitation.

\end{document}